\documentclass[11pt]{article}
\renewcommand \baselinestretch{1.3}
\usepackage[pdftex]{graphicx}

\usepackage{amsfonts,amsbsy,amsmath,amssymb}

\newcommand{\BE}{\begin{equation}}
\newcommand{\EE}{\end{equation}}
\newcommand{\BA}{\begin{eqnarray}}
\newcommand{\EA}{\end{eqnarray}}
\newcommand{\Pp}{{\scriptstyle{{\rm P}}}}

\newcommand{\msbar}{\overline{\rm MS}}
\newcommand{\alphQ}{\alpha_s(Q)}
\newcommand{\alphs}{\alpha_s}
\newcommand{\Lamt}{\tilde{\Lambda}}
\newcommand{\half}{\frac{1}{2}}

\begin{document}

\begin{titlepage}

\vspace*{20mm}
\begin{center}
              {\LARGE\bf  QCD Perturbation Theory: } \\
\vspace*{4mm}
             {\Large \bf It's not what you were taught}
\vspace{23mm}\\
{\large P. M. Stevenson}
\vspace{4mm}\\
{\it
T.W. Bonner Laboratory, Department of Physics and Astronomy,\\
Rice University, Houston, TX 77251, USA}

\vspace{33mm}

{\bf Abstract:}

\end{center}

\begin{quote}

    Physical quantities in QCD do not depend upon $\alphQ$.   There is no way to measure $\alphQ$ experimentally.
If those statements sound shocking, please read on.  They are actually well-known facts, though ones that are constantly being 
ignored in the QCD literature.  Renormalized perturbation theory is not an ordinary power-series expansion; its 
renormalization-scheme ambiguity is not merely a minor nuisance.  Rather, it is a structure in which invariance 
under redefinitions of the coupling is a fundamental symmetry --  a symmetry that, like any other, deserves 
respect.  I speak bluntly because without a change in mindset perturbative QCD can never become a proper, scientific enterprise.

\end{quote}

\end{titlepage}

\setcounter{equation}{0}

\section{Introduction}

   Suppose my abstract had begun: {\it  Physical quantities in QCD do not depend upon the gauge parameter $\xi(\mu)$.   There is no 
way to measure $\xi(\mu)$ experimentally.}  Would that not be boringly obvious?  Gauge invariance is a fundamental 
symmetry of QCD.  

   But there is another fundamental symmetry of the theory -- a symmetry of the renormalization process -- that somehow gets no respect.  
It is that of renormalization-scheme (RS) invariance.\footnote{
Also known as Renormalization-Group invariance \cite{SP,GML}, but see Appendix A.  
} 
The symmetry has a simple origin:  One may always make a change of variables without altering the physics.  Here, though, the change of 
variables involved -- from the bare to a renormalized coupling constant -- is from an infinitesimal quantity, ill-defined in the absence of regularization, 
to something finite.  For renormalization to make sense it is essential that the physics is independent of the arbitrary choices involved in defining 
the RS.  Without that symmetry the renormalization program would be a fraud.  Happily, though, it has been rigorously established, I believe.\footnote{
See Ref.~\cite{PopStora} and references therein.  
The result is referred to as ``the main theorem of perturbative renormalization theory.''
}
(The symmetry is quite explicitly manifest in the large-$N$ nonperturbative method \cite{RomLectures}.)

\begin{figure}[!h]
\centering
\includegraphics[width=0.8 \textwidth]{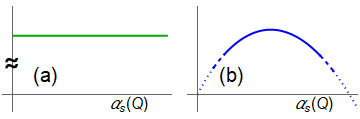}
\begin{quote}
\caption{{\small  A generic physical quantity as a function of $\alphQ$.  At left, the exact result.  At right, 
the next-to-leading order (NLO) approximation.  (``Unreasonable'' RS's are indicated by dashed/dotted lines.)    The value of the exact result is 
unknown, but its dependence upon $\alphQ$ -- viz. none at all -- is indisputable. }}
\label{RvsalphsFig}
\end{quote}
\vspace*{-5mm}
\end{figure}

    There is no problem for the theory itself, but there is a problem for practical theorists trying to make precise, numerical predictions for 
physically measurable quantities with only finite-order perturbative results.  The ``RS-dependence problem'' is that, while the {\it exact} 
predictions of QCD are exactly RS invariant -- they don't depend upon the definition (and hence can't depend upon the value) of  
``$\alphs$'' -- our perturbative {\it approximations}, at any finite order, {\it do} depend on it.   See Fig.~\ref{RvsalphsFig}.

\vspace{3mm}

\renewcommand \baselinestretch{1.0}

\renewcommand \baselinestretch{1.3}

     The prevailing attitude seems to be that this issue is merely a minor nuisance: ``{\it Just use any 
reasonable scheme.}''  It is that complacent attitude that I want to confront here.

\section{What is renormalized perturbation theory, actually?}

     ``Renormalized perturbation theory is a power-series expansion in powers of the renormalized 
coupling constant, $\alphQ$.''

     Yes, ... but what {\it is} ``$\alphs$'' actually?  What is its definition?  What is ``$Q$''?  
Without answers to these old questions (or to other, better questions) we are building on quicksand.  

    To define an ``$\alphs$'' one has to adopt a whole set of conventions -- a ``renormalization prescription'' (RP).\footnote{
I use the terminology that a {\it prescription} (RP) fixes everything in the {\it scheme} (RS) except for the value of the renormalization scale 
to be used. 
} 
The standard practice, adopted with little attempt at justification, is to use the so-called $\msbar$ prescription.  The two undoubted 
virtues of $\msbar$ are (i) it is convenient for the Feynman-diagram calculations and (ii) it is widely used, easing communication 
between different research groups.  However, it takes two chapters of a field-theory textbook to define and 
is obviously quite artificial.

   To define ``$Q$'' in general is impossible.  Supposedly one should use ``intuition'' (whose?) to identify the ``overall scale'' of 
the physical process being considered.  That is ambiguous even in the simplest cases, and even more so when there are multiple scales involved. 
How is one to find the ``right'' combination of kinematic variables?  

     The prescription and scale questions are deeply intertwined (which is why, viewed separately, they are not the right questions).  
They are two aspects of the wider RS-dependence problem.  The problem has been present since the beginning of quantum 
field theory, but its seriousness was first explicitly realized by Celmaster and Gonsalves \cite{CG}.  

     I do have something positive and specific to offer \cite{OPT}, explained in detail in a recent book \cite{PMSbook} 
(which also treats the case of factorized quantities, such as structure-function moments).  My aim here is not primarily to 
re-explain that method, but rather to attack the fallacy of saying: ``{\it Just use any reasonable scheme}.'' 

     The problem with that sentence is that it absolutely won't work without the word {\it reasonable}.  If one may use {\it any} scheme 
then one can obtain obviously ridiculous results (negative cross sections, for example). But if the notion of {\it reasonable} is so crucial, 
how do we define it?  Who is to decide?\footnote{
I am reminded of the distinguished judge who opined that, while he could not actually define ``obscenity,'' he knew it when he saw it.  That 
attitude, though humanly understandable, makes for poor jurisprudence -- and in science it is surely inadmissible.}

\section{Measuring the gauge parameter (a fable)}

     Imagine a world where physicists had discovered a theory to explain the strong interaction.  
They believed, rightly, that their theory was correct, but faced difficulty making precise predictions because -- and here
 is the key fictional premise -- their principal calculational  method produced results that depended upon the choice of gauge.  
They knew that their theory had to be gauge invariant.  Indeed, they could prove that their calculational method respected that 
invariance formally:  At any order of approximation the gauge dependence cancelled out in physical quantities, up to terms 
of higher order.  

      The mathematical physicists were content; the theory had been rigorously shown to be consistent:
Gauge fixing was necessary for concrete calculations, but the theory's predictions would be the same whatever 
gauge was used.    

     People concerned with obtaining experimentally testable predictions, however, had a problem:  
While the choice of gauge {\it  should not} matter, it nevertheless {\it did} matter in practice.   For a time the issue was debated, 
quite vigorously.  Various cures for the problem were suggested.  Most people, however, regarded the problem as just a minor 
nuisance. Shrugging their shoulders they adopted the mantra: ``{\it Just use any reasonable gauge}.''  

     A particular gauge choice, rather convenient for calculations, had originally been used by Prof. Emess, one of the great pioneers 
of the theory.  In practice it gave rather poor numerical results, but a fudge was quickly found by Prof. Barr that seemed to improve things.  
Gradually, this Emess-Barr gauge became the default choice.  It did not fix the gauge completely, however, and it was also necessary to 
fix $\mu$, an arbitrary scale parameter.  Supposedly this $\mu$ should be set equal to some kinematic variable $Q$ 
of the physical quantity in question.  The lack of an answer to the question {\it ``What is $Q$?''} did not seem to 
bother anyone much:  It was just a matter of using physical intuition to identify the ``typical momentum'' of the process.  

    In the early days of the theory this attitude worked well.  Most of the calculations were only leading order, or at best 
next-to-leading order (NLO), so one could not hope for too much precision.  There were a great many other issues to worry about 
(quark masses; weak and electromagnetic contributions; the need to parametrize and fit parton distribution and fragmentation 
functions; etc.) and a pressing need to get on calculating.  So the gauge-dependence problem drifted out of sight.  The conventional 
wisdom (use Emess-Barr with a reasonable $Q$) became so entrenched that authoritative reviews of the subject proudly 
presented detailed graphs, with experimental points and error bars, purporting to show the measured gauge parameter 
as a function of $Q$.  

    In vain did someone try to say:  ``Hang on a moment; the gauge parameter isn't a physical quantity!   What on earth does it mean 
to {\it measure} it?  In fact, isn't the theory supposed to be gauge invariant?  That is, the theory's predictions should not depend 
upon the gauge parameter {\it at all}.  The properties of the theory should be the same whatever value the gauge parameter is 
assigned.  So isn't a graph of the `measured gauge parameter' completely ridiculous?'' 

    In desperation to get his point across he turned to fiction, a little fable ... suppose the symmetry in question were not gauge 
invariance but Lorentz invariance ...

\section{It's a symmetry, dammit!}

    Now, the above story is fiction.\footnote{
In the real world, of course, our principal calculational method is perturbation theory, which respects gauge invariance {\it order-by-order} 
(provided we use a gauge-invariant RS) so that even {\it finite}-order results are {\it exactly} independent of the gauge choice.  
}
  However, everything in it would be true, I claim, if we replace gauge invariance with RS invariance.  Analogies are 
never perfect, but possible subtleties should not distract from the main point: In each case we are dealing with a 
{\it symmetry}, and symmetries deserve respect.  

     With gauge invariance we do not go about trying to find the ``right'' gauge for all of physics.  Nor do we try to characterize which gauges 
are ``reasonable'' and which aren't.  We simply use whatever gauge seems convenient for the calculation at hand, knowing that the results do not 
depend on that choice.  The same with Lorentz invariance; we do not try to identify the ``right'' frame of reference.  Yet much of the literature 
on scheme dependence has tried to identify the ``right'' RP and ``right'' scale $Q$ (with the latter dependent upon the context and 
the former assumed -- for no good reason  -- to be universal).    

     What, specifically, is the symmetry here?  It is this:\footnote{
As in the fable, there are a great many other issues to worry about.  To set these aside I focus here on an idealized QCD 
with a given, fixed number of species of massless quarks.  
}
 The renormalized coupling constant (couplant) $a \equiv \alpha_s/\pi $ has no unique definition and can be transformed 
to another one, $a'$, by transformations of the form
\BE
\label{ap}
a' = a(1+ v_1 a + v_2 a^2 + \ldots),
\EE
where the $v_i$ are finite but otherwise arbitrary.  The coefficients in the perturbation series of a physical 
observable,\footnote{
       The leading-order coefficient (which is RS invariant) has been scaled out for convenience.  More generally ${\cal R}$ could begin 
with $a^\Pp(1+ \ldots)$, but the generalization is entirely straightforward and brings in no new features so, for simplicity, I consider 
here only $\Pp=1$.  Only perturbative  quantities with an ordinary power-series form are being considered here.  For the case of 
factorized quantities, see Ref.~\cite{PMSbook}, Chap. 12.  
}
\BE
\label{Rser}
{\cal R} = a(1 +  r_1 a + r_2 a^2 + \ldots), 
\EE
then transform as 
\BA
{r_1}^\prime & = & r_1-v_1, \nonumber \\
{r_2}^\prime & = & r_2-2 r_1 v_1 + 2 v_1^2-v_2, \label{rp}  \\
& & \mbox{\rm etc.,} \nonumber
\EA
so that ${\cal R}$ remains invariant.  Because of this symmetry the coefficients $r_i$ and the couplant $a$ may take any values one likes 
(though it would be prudent to require $a>0$).  

    This symmetry arises naturally from the renormalization process.  If we calculate in terms of the {\it bare} couplant $a_B=g^2/(4 \pi^2)$, 
where $g$ is the gauge coupling present in the QCD Lagrangian, we find divergent integrals giving infinite coefficients in the expansion of 
${\cal R}$ in powers of $a_B$.  Renormalization, as far as physical quantities are concerned, is just a substitution 
\BE
\label{aBser}
a_B = a(1+ z_1 a + z_2 a^2 + \ldots),
\EE
where the $z_i$ have suitable infinite parts to give the necessary cancellations to yield finite coefficients $r_i$ in Eq.~(\ref{Rser}).  
(Regularization is needed to do this properly.)  To prove renormalizability -- as has been done, of course -- one has to show that there 
is {\it some} precise version of Eq.~(\ref{aBser}) for which this cancellation of infinities works to all orders.  
Other schemes just correspond to making different choices for the finite parts of the $z_i$'s, so that any two schemes are related 
by the transformation (\ref{ap}), with finite, but arbitrary, $v_i$'s.  The RS symmetry may now be discussed without further mention of 
regularization, which has been and gone, having done its job.  

     What makes this symmetry different from gauge invariance, or Lorentz invariance, is nothing to do with the theory itself, but is all
to do the the approximation method -- perturbation theory.  In any finite order the perturbative approximation to ${\cal R}$ is only invariant 
up to unknown terms of higher order.   That is not a problem for the theory, but it is a problem for the practical theorist.  

     To repeat, the problem is not with the {\it theory} but with the {\it approximation}.  It arises in {\it any} renormalizable theory.  
Fixating on the peculiarities of any particular theory, examining the entrails of its Feynman diagrams, is of no relevance.  What is needed 
is a clearer understanding of the symmetry, and more careful thought about what, precisely, a perturbative approximation actually is.  

\section{What are the invariants?}

      As with any symmetry, having specified its transformations, we should ask: {\it What are the invariants of this symmetry?}  
The physical quantities ${\cal R}$ are invariants, of course, but only in a formal sense.  Are there invariants that can actually be calculated 
in finite time?  Yes, there are \cite{OPT,PMSbook}.  

      First, I need to review the concepts of the ``$\beta$ function'' and ``Dimensional Transmutation.''  Any RS necessarily 
involves a renormalization scale, $\mu$.  In traditional momentum-subtraction schemes it corresponds to the 
momentum entering some 3-point vertex function.  In $\msbar$ it arises rather differently, but all that matters here is the fact that 
one parameter of the RS must have dimensions of mass.  The couplant depends upon this $\mu$.   That is surprising because Dimensional 
Analysis tells us that, in a theory with no dimensionful parameters, it is impossible to write down a unique function $a(\mu)$ 
that is not constant.  The resolution of that paradox is that $a(\mu)$ is not given {\it uniquely}.  Rather it is given by a differential equation
\BE
 \mu \frac{d a}{d \mu} \equiv \beta(a) = -b a^2(1+ c a + c_2 a^2 + \ldots),  
\EE
with a missing boundary condition, reflecting the fact the Lagrangian was not unique, but had a free parameter, the bare $g$.  
Na\"ively $g$ is dimensionless, but as it's not finite we can't express results in terms of it.  (The issue of how to properly 
parametrize the one-parameter ambiguity will be addressed in Appendix B.)   

     The $\beta$ function depends upon the RP.  It transforms as:
\BE
  \beta'(a') \equiv \mu \frac{d a'}{d \mu} =  \mu \frac{d a}{d \mu}  \frac{d a'}{d a} = \beta(a) \frac{d a'}{d a}.
\EE
Thus, in a different RP we would have 
\BE
\beta'(a') = -b a'^2(1+ c a' + c_2' a'^2 + \ldots),
\EE
where, remarkably, the $c$ coefficient \cite{tH}, as well as $b$, is the same.  The higher coefficients, though, are not the same.  In particular
\BE
\label{c2p}
{c_2}^\prime = c_2 + v_2 - v_1^2-v_1 c.
\EE
(See Ref.~\cite{PMSbook} for an explicit presentation of the simple algebra involved.)  

    One can now show that there are calculable invariants that are combinations of the $r_i$ coefficients and the $c_j$ coefficients 
of the $\beta$ function. The first is just $c$.  The second is 
\BE
\label{rho2}
\rho_2 \equiv  c_2 + r_2 - c r_1 - r_1^2.  
\EE
The reader is invited to check this RS invariance explicitly, by substituting in $\rho_2 \equiv  c_2' + r_2' - c r_1' - r_1'^2$ using Eqs. 
(\ref{rp}, \ref{c2p}) and seeing the cancellation of all $v_1$ and $v_2$ terms.  Having done that, it should be easy to believe that 
there are further invariants in higher orders.  The next is 
\BE
\rho_3 = c_3 - 2 c_2 r_1 + c r_1^2 + 4 r_1^3-6 r_1 r_2 + 2 r_3,
\EE
and one can give implicit and explicit formulas for the general case \cite{PMSbook}.  

      There is one more invariant quantity, present even at NLO.   That invariant is
\BE
\label{rho1Qdef}
{\boldsymbol \rho}_1(Q)  \equiv b \ln(\mu/\tilde{\Lambda}) - r_1.  
\EE
Uniquely, it depends upon a kinematic scale $Q$ associated with the physical quantity ${\cal R}$ that enters in the calculated $r_1$ 
coefficient, which has necessarily has the form 
\BE
\label{r1form1}
r_1= b \ln(\mu/Q) + r_{1,o}.
\EE
 (Using a different ``$Q$'' would just move a $b \ln(Q_{\rm new}/Q)$ term between the two terms of this decomposition and so does not 
affect the value of ${\boldsymbol \rho}_1(Q)$.)  
Clearly ${\boldsymbol \rho}_1(Q)$ is independent of $\mu$, but equally importantly it is independent of RP.  That fact is shown in Appendix B, 
which explains the definition of the $\Lamt$ parameter and its RP dependence.  

     As with any symmetry, the invariants are of paramount importance.  Any physically meaningful result for, or approximation to, ${\cal R}$ 
 must be expressible in terms of invariants.

\section{What is the perturbative approximation?}

    We do need to be specific about what the perturbative approximation, at any given order, actually is.  If that is left ambiguous 
then there is no use trying to settle the RS issue.   

    The first thing to say is that we do not want a truncated expansion in $1/\!\ln (Q/\Lambda)$.   While appropriate for  
discussing the $Q \to \infty$ limit, it is not a satisfactory basis for an approximation method at finite $Q$.  
There are two main problems:  (i) It produces two artificial and unnecessary ambiguities; {\it What is $Q$?} and {\it How is $\Lambda$ defined?}.  
(ii) It is manifestly inferior to an expansion in powers of the couplant, since it breaks down for $Q \le \Lambda$, whereas the couplant can 
(and beyond NLO {\it does},  in sensible schemes) remain finite at all $Q$.  An approximation that gives an absurd answer at $Q=\Lamt$ can hardly 
be expected to give a good approximation when $Q$ is a little bit larger and, quite likely, will be inferior at all $Q$.  

     So, our perturbative approximation should be directly based on the perturbation expansion in powers of $a$.  However, we need to 
recognize that we must truncate {\it both} the series for ${\cal R}$ and that for the $\beta$ function .  We need a well-defined approximation 
to $\beta(a)$ because it is needed to express the result in terms of (our choice of) the free parameter of the theory.  It is natural to truncate 
both series after same number of terms because what matters is how many of the invariants we know.  Thus, the  N$^k$LO approximation 
should be defined as 
\BE
{\cal R}^{(k+1)} = a(1+r_1 a + \ldots + r_k a^k),
\EE
where $a$ here is short for $a^{(k+1)}$, the solution to the integrated $\beta$-function equation with $\beta$ replaced by $\beta^{(k+1)}$:
\BE
\beta^{(k+1)} \equiv - b a^2(1+ c a + \ldots + c_k a^k).
\EE

The approximant ${\cal R}^{(k+1)}$ is a function of the RS parameters $b \ln(\mu/\Lamt),c_2,\ldots,c_k$ and the invariants 
${\boldsymbol \rho}_1(Q), c, \rho_2,\ldots,\rho_k$.  It is convenient to use the integrated $\beta$-function equation to swap 
$\mu/\Lamt$ for $a$ itself, so that the RS parameters are now $a, c_2, \ldots,c_k$.   For example, in NLO one just needs to rearrange 
Eq.~(\ref{rho1Qdef}) to find $r_1$ and then substitute in ${\cal R}^{(2)} = a(1+r_1 a)$ to show that
\BE
\label{R2fun}
{\cal R}^{(2)} = a \left( 2 + c a \ln \left| \frac{c a}{1+ c a} \right| - {\boldsymbol \rho}_1(Q) a \right).
\EE
One can then plot ${\cal R}^{(2)}$ explicitly as a function of $a \equiv \alphs/\pi$, as in Fig.~\ref{RvsalphsFig}(b).

\section{Respect the symmetry, resolve the problem}

    Having emphasized the inadequacy of the prevailing viewpoint, and reviewed some salient facts, I now come to the question of how to 
resolve the problem.  

     The key is a very simple point:  Since the exact result is exactly independent of the RS, it would not be reasonable for the chosen 
approximate result to be very sensitive to small changes of RS.  So perhaps the ``most reasonable'' RS is that for which the approximation 
is {\it  stationary} under small RS changes.  The approximation then manifests a local symmetry that mimics the exact symmetry of the 
exact result.  This is the {\it Principle of Minimal Sensitivity} \cite{OPT,PMSbook}.  

      It  is not a theorem.  It is a matter of opinion, perhaps, but it is supported by plentiful evidence -- because, as always, we can study 
simpler examples and soluble cases to see if, how, and why an approximation method works.  There are actually many situations akin to the 
RS-dependence problem, where the exact result is invariant  -- independent of certain ``extraneous variables'' -- while the 
approximate result is not.  I would urge readers to spend a little time studying such examples; see Chaps 4, 5 of Ref.~\cite{PMSbook}.  
The idea has been used successfully in other fields; for example in Ref.~\cite{ChenHaule}.  

\begin{figure}[!hbt]
\centering
\includegraphics[width=0.7 \textwidth]{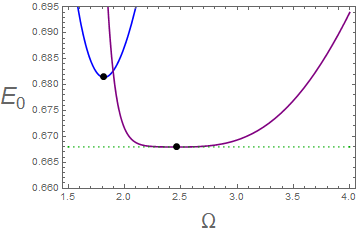}
\begin{quote}
\caption{{\small   Results for $E_0$ in first and fifth orders of the CK expansion, as a function of the extraneous variable $\Omega$.  
The flat region grows and moves to the right.  For any fixed $\Omega$ the results diverge, but the ``optimized'' results converge nicely.  
(Adapted from Ref.~\cite{PMSbook}, Fig.~5.2.)}}
\label{PMSQOhigh}
\end{quote}
\vspace*{-5mm}
\end{figure}

       A particularly instructive example is the quartic oscillator in quantum mechanics in the Caswell-Killingbeck (CK) \cite{CK} 
(or linear $\delta$) expansion, where one adds and subtracts a $\half \Omega^2 x^2$ term to the Hamiltonian and uses an oscillator 
of frequency $\Omega$ as the $H_0$.  Fig.~\ref{PMSQOhigh} shows results for the ground-state energy, $E_0$, as a function of the 
extraneous parameter $\Omega$.  Here the ``minimal sensitivity'' argument\footnote{
made independently by both Caswell and Killingbeck.
}
clearly yields good results, and for exactly the advertised reason.  The exact result is completely independent 
of $\Omega$, and where the approximation is flat in $\Omega$ is where it works best.  As the order 
increases the flat region moves to the right, while growing in extent.  If one sits at any fixed $\Omega$ value the results diverge, 
but if we choose $\Omega$ by minimal sensitivity in each order then we stay with the flat region and the results converge.

\vspace{5mm}

\renewcommand \baselinestretch{1.0}

\renewcommand \baselinestretch{1.3}

\begin{figure}[!h]
\centering
\includegraphics[width=0.65 \textwidth]{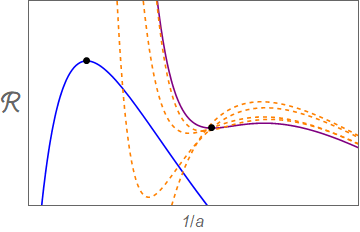}
\begin{quote}
\caption{{\small  Sketches of perturbative approximations to a generic physical quantity as functions of $1/a$.  At left,  NLO; at right, a much 
higher-order approximation, with the solid curve being for the optimized values of the $c_2, c_3, \ldots$ parameters, and the dashed curves 
illustrating the effect of varying those parameters.  The black dot indicates the ``optimized'' result in each case.   (As with the previous figure, 
the vertical extent is about twice the error estimate on the NLO result.)  The flat region grows and shifts rightward:  Thus, what was a good 
value for $\alphs$ in NLO is not at all a good value to use in higher orders.}}
\label{R4}
\end{quote}
\vspace*{-5mm}
\end{figure}

   In QCD, in the $R_{e^+e^-}$ case, we can see the same sort of pattern emerging, especially if we plot against $1/a$ rather than $a$ itself.  
Fig.~\ref{R4} shows sketches of the NLO and a much higher-order approximation as a function of $1/a$  and the other RS variables.\footnote{
The ``higher order'' curves are fiction, but are based on actual results in N$^3$LO, shifted more to the right, in accordance with the 
consistent trend seen in the lower-order results. See Figs. 8.2, 8.3 in Ref.~\cite{PMSbook}.  
}
(Here there are more and more scheme variables as the order increases.)  Again the trend is for the flat region to grow and shift to the right.  
Note that  a ``reasonable'' value for $\alphs$ in NLO is not at all a ``reasonable'' value to use in much higher orders. The optimal $\alphs$ value 
evolves with order, and that is essential for convergence.

\vspace{5mm}

\renewcommand \baselinestretch{1.0}

\renewcommand \baselinestretch{1.3}

\section{Final remarks}

      Any finite-order approximation comes with uncertainties.  That is not, though, an excuse for tolerating unquantifiable, 
open-ended ambiguities (that will only get worse at higher orders) arising from vague and faulty concepts.  
 The ``optimization'' method, at any given order, yields a definite, reproducible result that does not depend at all upon the 
RS used in the Feynman-diagram calculations, because it only depends on the RS invariants 
${\boldsymbol \rho}_1(Q),c,\rho_2,\rho_3,\ldots$.  The error can be estimated in the usual way by examining the apparent 
convergence of the calculated terms of the ${\cal R}$ series in the optimal scheme.  The optimal couplant shrinks with order, 
producing results that converge and become flatter and flatter in the RS variables.

   Physical quantities themselves do not depend on $\alphs$.    Perturbative approximations do.  For a given physical quantity, 
at a given energy and at a given order, there is an ``optimal'' $\alphs$ (along with optimal values of the other RS parameters 
$c_2,\ldots,c_k$) that produces the most believable result for that case.  One might put it like this:  ``$\alphs$'' is not a 
{\it thing}; it is a {\it tool}; a shape-shifting tool that, when used properly, will adjust itself to suit the task at hand.  If we think we have 
measured $\alphQ$, based on low-order calculations, we will come badly unstuck when we try to improve our precision by calculating higher orders.   

     Here I have discussed only purely perturbative physical quantities, but many QCD predictions involve parton distribution and 
fragmentation functions that are nonperturbative and have to be taken from experiment.  The factorization into perturbative 
and nonperturbative parts involves factorization-scheme ambiguities, which are entangled with RS ambiguities.  Chap.~12 of 
Ref.~\cite{PMSbook} treats this problem in the moments formalism, building on earlier work of other authors \cite{Pol,NN}.  
A key result, realized first by Nakkagawa and Ni\'egawa \cite{NN}, is that ``optimization'' results in the coefficient function being unity, 
with all perturbative corrections effectively exponentiated into the anomalous-dimension term.  What is needed is to translate those results 
into the language of distribution functions and parton-evolution (DGLAP) equations.  QCD phenomenology can then be put on a firm foundation.  

   The obstacle to scientific progress, it has been wisely said, is often not the difficulty of entertaining new ideas, but of letting go of old ideas 
that have outlived their usefulness.  It took decades for the assumption that the coupling constant was a {\it fixed} thing to be supplanted by the 
notion of an {\it effective} or {\it running} coupling constant, $\alphQ$.  That idea, renormalization-group-improved perturbation theory, was a great 
advance, but it was never unambiguous, having no answers to the questions {\it What is $Q$?} and {\it Which RP is to be used?}.  It is time 
it was pensioned off.

\newpage
\vspace{-10mm}

\section*{Appendix A:  The ``Renormalization Group'' }

       The symmetry of RS invariance is also known as ``Renormalization Group (RG) invariance.''   That can perhaps create confusion, 
because the ``RG'' label has now come to mean so many different things.  An excellent account of the early history presented by 
Fraser \cite{Fraser} is most illuminating.  Modern use of the ``RG'' label covers a wide spectrum of powerful ideas, applicable to 
all sorts of theories, revolving around transformations of scale.  However, the {\it groupe de normalization} introduced by Stueckelberg and 
Peterman (SP) \cite{SP} is specific to renormalizable theories and involves many parameters, one more in each order.  Scale dependence 
($\mu/\Lamt$ dependence) is the most important aspect, entering at the lowest order, but the other aspects are on an equal footing.  
When I pointed out that the $\beta$-function coefficients $c_2, c_3, \ldots$ could be used as RS parameters \cite{OPT} it was not an 
``extension'' of SP's {\it groupe} but just a concrete realization of it.  

        Is the RG really a group?   For the modern usage, the question may not be meaningful, but for what SP and I are talking about 
there is certainly a group.  The transformations $a'=a(1+v_1 a+ \ldots)$ are invertible; two successive transformations are equivalent 
to a single transformation of the same kind; there is obviously an identity transformation.  There is, though, nothing very interesting 
in the group-theory properties.  The transformations do not commute, so one might think that there is fun to be had with non-Abelian 
structure constants, etc..  However, the fact that RS's can be parametrized by $\tau, c_2, c_3,\ldots$, means that one can describe 
the group in terms of translations $\tau \to \tau'$, $c_2 \to c_2'$, etc..  The explicit connection to the $a \to a'$ transformation is easily 
worked out;  see Exercise 7.4 of Ref.~\cite{PMSbook}.

\section*{Appendix B: The $\Lamt$ parameter and the ${\boldsymbol \rho}_1(Q)$ invariant}

\label{AppLam}

    Integration of the $\beta$-function equation leads to 
\BE
     \ln \mu = \int \! \frac{da}{\beta(a)} + {\mbox{\rm const.}}
\EE
and the free parameter of QCD enters when fixing the constant of integration.  Trying to set $a(\mu_0)$ to a fixed value at some finite 
scale $\mu_0$ is not satisfactory.\footnote{
If two people use definitions with  different scales and prescriptions, how are we to to inter-convert between them?  Do we evolve the scale 
in the first scheme or the second?  Do we convert using $a'=a(1+v_1 a+\ldots)$ or by its inverse $a=a'(1-v_1 a'+\ldots)$, and to what order 
do we truncate those series?
}
The basic problem is that finite energies are not entirely under perturbative control.   In any conventional RP the $\beta$-function series will 
be {\it divergent}, so the value of $a(\mu_0)$ will, at sufficiently high orders, tend to vary wildly from one order to the next: 
 Forcing it to be a fixed value will destabilize the ultraviolet (uv) behaviour, which is what should be kept fixed.  

    The uv behaviour of $a(\mu)$ {\it is} under full perturbative control.   The leading asymptotic behaviour, of course, is $1/a=b \ln \mu + O(1)$, 
where it does not matter what units $\mu$ is measured in.  The free parameter of QCD, a scale $\Lamt$, enters when we consider the 
next-to-leading asymptotics.  In a theory with $c=0$ that would be just $1/a = b \ln(\mu/\Lamt) + O(1/\ln \mu)$.  Allowing for non-zero 
$c$ the NLO form is
\BE
\label{NLOa}
\frac{1}{a} = \tau + c \ln |\tau/c| + O\left( \frac{\ln \tau}{\tau} \right),
\EE
where
\BE
\tau \equiv b \ln(\mu/\Lamt).
\EE
(The tilde is used to distinguish this $\Lambda$ definition from that of Buras {\it et al} \cite{BBDM}.)  

      It is easy to show that Eq.~(\ref{NLOa}) results from defining $\Lamt$ by 
\BE
\label{lamdef}
 \ln(\mu/\Lamt) = \lim_{\delta \to 0} \left( \int_\delta^a \frac{dx}{\beta(x)} + {\cal C}(\delta) \right), 
\EE
with 
\BE
\label{Cdeltadef}
{\cal C}(\delta) \equiv \int_\delta^\infty \frac{dx}{b x^2(1+c x)}.
\EE
That is, the $\beta$-function equation is integrated using the natural perturbative range, $0$ to $a$, with a 
tiny bit of mathematical sophistication needed to deal with the divergence of the integral at the $x=0$ endpoint.  
Any valid $\Lambda$ definition will be equivalent to Eq.~(\ref{lamdef}) up to the addition of a finite, 
RS-invariant constant to the right-hand side.  

      The $\Lamt$ parameter just defined depends upon the RP being used.  However, that RP dependence involves 
{\it only} the $v_1$ parameter of Eq.~(\ref{ap}).  The $\Lamt$ parameters of two RP's 
are related exactly by 
\BE
\label{CGrel}
\ln(\Lamt'/\Lamt) = v_1/b.  
\EE
This crucial result was proved by Celmaster and Gonsalves (CG) \cite{CG}.  For alternative proofs, see \cite{PMSbook,MaxConDNW}.  

    Thanks to the CG relation, the ${\boldsymbol \rho}_1(Q)$ quantity, Eq.~(\ref{rho1Qdef}), is independent of both $\mu$ and the RP. 
In fact, the decomposition of RS dependence into separate scale and prescription dependences, though necessary pedagogically, is not 
very meaningful mathematically.  The variable that matters at NLO is not $\mu$ itself but the {\it ratio} of $\mu$ to $\Lamt$.  At higher orders 
one can fully characterize the RS dependence of ${\cal R}^{(k+1)}$ by the variables $\tau$, $c_2,\ldots,c_k$ \cite{OPT,PMSbook}.

    The $\Lamt$ parameter is directly related to the uv behaviour of physical quantities as $Q \to \infty$.   Eq.~(\ref{NLOa}) gives $a$ 
as a series in $\frac{1}{\tau}$, and one may use ${\boldsymbol \rho}_1(Q) = \tau - r_1$ to express $a$, and hence 
${\cal R}=a(1+r_1 a + \ldots)$, as a series in $1/{\boldsymbol \rho}_1(Q)$.  The intermediate expansion in $1/\tau$ is justified 
provided $\mu$ is chosen so that $\mu \propto Q$ asymptotically.  The RS-invariant asymptotic result is 
\BE 
\frac{1}{{\cal R}} =  {\boldsymbol \rho}_1(Q) + c \ln | {\boldsymbol \rho}_1(Q)/c| + 
O\left( \frac{ \ln {\boldsymbol \rho}_1(Q)}{ {\boldsymbol \rho}_1(Q)} \right).
\EE
Note that one may write 
\BE
\label{LamR}
{\boldsymbol \rho}_1(Q) = b \ln(Q/\Lamt_{\cal R}),
\EE
where the  $\Lamt_{\cal R}$ parameter is related to the $\Lamt$ parameter of the RS used to calculate $r_{1,o}$ by
\BE
\ln(\Lamt_{\cal R}/\Lamt) = r_{1,o}/b,
\EE
which can be viewed as a particular case of Eq.~(\ref{CGrel}).

       So, what is ``the free parameter'' of QCD?  There is no unique choice -- reflecting the indirect entry of a scale 
into a seemingly scale-invariant theory.  There are two choices involved: the definition of $\Lambda$ in any given RP, and the 
choice of a ``reference'' RP.  Different people may reasonably make different choices.  However, the conversion between any two 
choices can be made {\it exactly} -- and that is true in practical terms, not just in principle.  It is no different from the conversion 
between two, equally well-defined, systems of units; a bit of a nuisance, but in no sense a problem.

\end{document}